\pdfoutput=1
\documentclass[11pt,twoside,a4paper,cmspaper,final,collab]{cms-tdr}
\def\svnVersion{ef95314}\def\svnDate{2021/11/03}\def\cmsCernNoTag{CERN-EP-2021-111}\def\cmsCernDate{\today}\def\cmsMessage{Published in Physical Review Letters as \href{http://dx.doi.org/10.1103/PhysRevLett.127.191801}{\doi{10.1103/PhysRevLett.127.191801}.}}
\begin{document}\cmsNoteHeader{SMP-20-012}

\newcommand{\WW}{\ensuremath{\PW \PW}\xspace}
\newcommand{\WZ}{\ensuremath{\PW \PZ}\xspace}
\newcommand{\ZZ}{\ensuremath{\PZ \PZ}\xspace}
\newcommand{\MATRIX}{\textsc{matrix}}
\newcommand{\lepW}{\ensuremath{\ell_{\PW}}\xspace}
\newcommand{\lepZI}{\ensuremath{\ell_{\PZ}}\xspace}
\newcommand{\lepZII}{\ensuremath{\ell'_{\PZ}}\xspace}

\newlength\cmsFigWidth
\newlength\cmsTabSkip\setlength{\cmsTabSkip}{1ex}
\ifthenelse{\boolean{cms@external}}{\providecommand{\cmsTable}[1]{#1}}{\providecommand{\cmsTable}[1]{\resizebox{\textwidth}{!}{#1}}}
\ifthenelse{\boolean{cms@external}}{\setlength\cmsFigWidth{0.49\textwidth}}{\setlength\cmsFigWidth{0.65\textwidth}} 
\ifthenelse{\boolean{cms@external}}{\providecommand{\cmsLeft}{upper\xspace}}{\providecommand{\cmsLeft}{left\xspace}}
\ifthenelse{\boolean{cms@external}}{\providecommand{\cmsRight}{lower\xspace}}{\providecommand{\cmsRight}{right\xspace}}

\cmsNoteHeader{SMP-20-012} 
\title{Measurements of the electroweak diboson production cross sections in proton-proton collisions at \texorpdfstring{$\sqrt{s} = 5.02\TeV$}{sqrt(s) = 5.02 TeV} using leptonic decays }

\date{\today}

\abstract{
The first measurements of diboson production cross sections in proton-proton interactions at a center-of-mass energy of 5.02\TeV are reported. They are based on data collected with the CMS detector at the LHC, corresponding to an integrated luminosity of 302\pbinv. Events with two, three, or four charged light leptons (electrons or muons) in the final state are analyzed. The \WW, \WZ, and \ZZ total cross sections are measured as $\sigma_{\WW}=37.0^{+5.5}_{-5.2}\stat^{+2.7}_{-2.6}\syst\unit{pb}$, $\sigma_{\WZ}=6.4^{+2.5}_{-2.1}\stat^{+0.5}_{-0.3}\syst\unit{pb}$, and $\sigma_{\ZZ}=5.3^{+2.5}_{-2.1}\stat^{+0.5}_{-0.4}\syst\unit{pb}$. All measurements are in good agreement with theoretical calculations at combined next-to-next-to-leading order quantum chromodynamics and next-to-leading order electroweak accuracy.
}

\hypersetup{
pdfauthor={CMS Collaboration},
pdftitle={Measurements of the electroweak diboson production cross sections in proton-proton collisions at sqrt{s}= 5.02 TeV using leptonic decays},
pdfsubject={CMS},
pdfkeywords={CMS, SM}}

\maketitle 

The study of diboson production at the CERN LHC is an important test of the standard model (SM) of particle physics
because of its sensitivity to the self-interactions between gauge bosons via trilinear gauge couplings~\cite{Hagiwara:1993ck}. Understanding 
diboson production is also important for Higgs boson measurements and for a multitude of beyond-the-SM searches
where these diboson processes represent irreducible background contributions. The CMS and ATLAS Collaborations have 
measured diboson production cross sections in proton-proton ($\Pp\Pp$) collisions at center-of-mass energies of 7, 8, and 
13\TeV~\cite{ATLAS:2012mec, Aad:2016wpd, Aaboud:2017qkn, Aad:2012twa, Aad:2016ett, Aaboud:2019gxl, Aad:2012awa, 
Aaboud:2016urj, Aaboud:2017rwm, Chatrchyan:2013yaa, Khachatryan:2015sga, Sirunyan:2020jtq, Khachatryan:2016poo, Sirunyan:2019bez, 
Chatrchyan:2012sga, CMS:2014xja, Sirunyan:2020pub}. In this Letter, we present the first measurements of 
electroweak diboson production cross sections at $\sqrt{s}=5.02\TeV$. All measurements are performed with $\Pp\Pp$ collision data 
corresponding to an integrated luminosity of 302\pbinv, collected in November 2017 with the CMS detector~\cite{Chatrchyan:2008zzk} 
at the LHC. The maximum instantaneous luminosity delivered by the LHC during this period was $1.37 
\times 10^{33} \percms$, and the mean number of $\Pp\Pp$ interactions per bunch crossing, assuming a total inelastic cross section of 65\unit{mb}, was 2.0. 

The central feature of the CMS apparatus is a superconducting solenoid of 6\unit{m} internal diameter, 
providing a magnetic field of 3.8\unit{T}. Within the solenoid volume are a silicon pixel and strip 
tracker, a lead tungstate crystal electromagnetic calorimeter (ECAL), and a brass and scintillator 
hadron calorimeter (HCAL), each composed of a barrel and two end cap sections. Forward calorimeters extend 
the pseudorapidity ($\eta$) coverage provided by the barrel and end cap detectors. Muons are detected in 
gas-ionization chambers embedded in the steel flux-return yoke outside the solenoid. A more detailed 
description of the CMS detector, together with a definition of the coordinate system used and the 
relevant kinematic variables, is reported in Ref.~\cite{Chatrchyan:2008zzk}. 

Signal and background processes are simulated by using several Monte Carlo (MC) generators. The propagation of the 
generated particles through the CMS detector and the modeling of the detector response is performed using 
\GEANTfour~\cite{GEANT4}, assuming alignment and calibration from real data.
Simulated signal events are generated at next-to-leading order (NLO) in perturbative quantum chromodynamics (QCD) 
and dynamic renormalization and factorization scales using \POWHEG (v2)~\cite{Nason:2004rx,Frixione:2007vw,Alioli:2010xd}. 
The \WW, \WZ, and \ZZ signal cross sections are scaled from NLO to next-to-next-to-leading order (NNLO) using \MATRIX\ 
calculations~\cite{Grazzini:2017mhc}. The \MGvATNLO~\cite{Alwall:2014hca} generator is 
used to simulate \PW and \PZ/$\Pgg^*$+jets at NLO. The simulation includes up to two extra partons at the matrix element 
level and uses the FxFx merging scheme~\cite{Frederix2012}.  
Simulated top quark events---top quark pair production (\ttbar) and single top quark processes---are also generated using \POWHEG.
All the events are then interfaced with \PYTHIA8 (v8.2)~\cite{Sjostrand:2014zea} for parton showering, hadronization, 
and the underlying event simulation, using the CP5 tune~\cite{Skands,CP5}. The NNPDF3.1~\cite{NNPDF3.1} 
NNLO parton distribution functions (PDFs) are used. The simulated samples include pileup collisions.

The particle-flow (PF) algorithm~\cite{CMS-PRF-14-001} reconstructs and identifies each individual particle in an 
event, with an optimized combination of information from the various elements of the CMS detector. The energy of photons 
is obtained from the ECAL measurement. The energy of electrons is determined from a combination of the electron momentum 
at the primary interaction vertex as determined by the tracker, the energy of the corresponding ECAL cluster, and the 
energy sum of all bremsstrahlung photons spatially compatible with originating from the electron track. The energy of 
muons is obtained from the curvature of the corresponding track. The energy of charged hadrons is determined from a 
combination of their momentum measured in the tracker and the matching ECAL and HCAL energy deposits, corrected for the 
response function of the calorimeters to hadronic showers. Finally, the energy of neutral hadrons is obtained from the 
corresponding corrected ECAL and HCAL energies. For each event, hadronic jets are clustered from these reconstructed 
particles using the anti-\kt algorithm~\cite{Cacciari:2008gp, Cacciari:2011ma} with a distance parameter of 0.4. Jet 
momentum is determined as the vectorial sum of all particle momenta in the jet, and, based on simulation, is typically 
within 5 to 10\% of the true momentum over the entire transverse momentum (\pt) spectrum and detector acceptance.
The missing transverse momentum vector \ptvecmiss is computed as the negative vector \pt sum of all the PF 
candidates in an event, and its magnitude is denoted as \ptmiss~\cite{Sirunyan:2019kia}. The \ptvecmiss 
is modified to account for corrections to the energy scale of the reconstructed jets in the event~\cite{Khachatryan:2016kdb}. 
The candidate vertex with the largest value of summed physics-object $\pt^2$ is the primary $\Pp\Pp$ interaction 
vertex. The physics objects used to calculate the primary vertex are the leptons, jets and the \ptmiss associated with this event.

Electrons are identified with a multivariate analysis (MVA) discriminant~\cite{Khachatryan:2015hwa,Sirunyan:2020ycc} and are required to have a 
\pt larger than 8\GeV and $\abs{\eta}<2.5$. Electrons matched to a secondary vertex consistent with a photon conversion or 
having at least one missing hit in the pixel tracking system are vetoed. Muons are identified as tracks in the central tracker 
consistent with either a track or several hits in the muon system, and associated with calorimeter deposits compatible with the 
muon hypothesis. Reconstructed muons are required to have $\pt>8\GeV$ and $\abs{\eta}<2.4$, and must fulfill criteria on the 
geometrical matching between the tracker and the muon track, and the quality of the global fit~\cite{Sirunyan:2018}. An upper 
threshold of 0.4 on the relative isolation (as defined in Ref.~\cite{Khachatryan:2016kod}), is applied for both electrons and muons. Lepton candidates are 
selected if the transverse (longitudinal) impact parameter with respect to the primary vertex does not exceed 0.05 (0.1)\unit{cm}. 
Identification criteria are specifically designed to separate prompt leptons---electrons and muons from the decay of a \PW or \PZ 
boson either directly or mediated by a leptonic \Pgt decay---from leptons that arise from other sources, 
such as \PQc or \PQb decays.
Two lepton categories, loose and tight, are defined. The loose identification criteria 
refer to the requirements presented above and are used to provisionally select all leptons in the events. The tight selection is based 
on an MVA discriminant---a gradient boosted decision tree---trained to separate between prompt and nonprompt lepton 
sources~\cite{Sirunyan:2020icl} and includes, in addition to the loose selection, (i) a tighter upper threshold on the relative isolation (0.085
for electrons and 0.325 for electrons and muons), and (ii) a threshold on the \PQb-tagging \textsc{DeepJet} discriminator~\cite{btagging,
Bols:2020bkb,CMS-DP-2018-058} for any jet that contains a lepton to reduce the \ttbar background.
Jets with $\pt>25\GeV$ and $\abs{\eta}<2.4$ which do not overlap ($\Delta R<0.4$) with a lepton passing the loose criteria 
are selected. 

Events of interest are selected using a two-tiered trigger system~\cite{Khachatryan:2016bia}. The first level (L1), composed 
of custom hardware processors, uses information from the calorimeters and muon detectors to select events at a rate of around 
100\unit{kHz}.  
The second level, known as the high-level trigger, consists of a 
farm of processors running a version of the full event reconstruction software optimized for fast processing, and reduces the 
event rate to around 1\unit{kHz} before data storage. Events that pass at least one single-lepton trigger  with \pt thresholds 
of 17 (electrons) or 12 (muons)\GeV are selected. Candidate events are further required to 
have at least two loose leptons (electrons or muons, $\ell$) with a minimal invariant mass of any lepton pair greater than 12\GeV.

The \WW signal region (SR) requirements are exactly two tight leptons with opposite charge and different flavor; the \pt of the 
leading (subleading) lepton $>$20 (10)\GeV; the \pt of the dilepton 
system $>$20\GeV; the azimuthal separation between the two leptons $>$2.8\unit{radians}; and the
transverse mass of any combination of a lepton with the \ptvecmiss $>$20\GeV. In addition, events with jets are rejected.

For the \WZ measurement two SRs are defined: one with three leptons (3$\ell$) and another with two muons with the same electric 
charge (2$\mu$ss). In the 3$\ell$ category, events with exactly three loose leptons with at least one opposite-sign 
same-flavor (OSSF) pair are selected. To exploit the characteristic kinematics of on-shell \WZ production, an  
algorithm is applied to tag the two leptons from the \PZ boson decay (\lepZI\ and \lepZII) and that of the \PW boson decay (\lepW). 
If only one OSSF lepton pair occurs in the event, the leptons corresponding to it are tagged as \lepZI\ and \lepZII, whereas the 
different flavor one is tagged as \lepW. If multiple OSSF pairs are found, the OSSF lepton pair with invariant mass closest to 
that of the \PZ boson is selected for the \lepZI \lepZII\ pair. Then, a \pt threshold of 8\GeV is imposed. Additional 
selection criteria are applied to increase the purity of \WZ events. The invariant mass of the \lepZI\ and \lepZII\ 
lepton pair must be consistent with the \PZ boson mass, $\abs{m_{\lepZI \lepZII} - 91.2\GeV} <$ 30\GeV. The two same-sign leptons are required
to pass the tight lepton requirements, the \pt of \lepW\ must be $>$20\GeV, and the invariant mass of the three lepton 
system must be $>$100\GeV. For the 2$\mu$ss category, events with two tight muons and zero jets are selected. 
The \pt of the leading (subleading) muon must be $>$20~(10)\GeV. To ensure a high-quality charge measurement, 
the relative uncertainty in the curvature of the muon track must be $<$20\%. Additionally, a minimal requirement of $\ptmiss > 25\GeV$ is included. 

For the \ZZ measurement, two categories are defined: one with four leptons (4$\ell$) and another with two leptons (2$\ell$2$\nu$). 
For the 4$\ell$ category, exactly four loose leptons with $\pt > 8\GeV$ are required. For the 2$\ell$2$\nu$ category, events 
with exactly two tight OSSF leptons are selected. The \pt of the leading (subleading) lepton must be $>$20~(10)\GeV. 
The invariant mass of the leptons must be close to the \PZ boson peak, $\abs{m_{\ell, \ell'} - 91.2\GeV}<10\GeV$. The axial \ptmiss 
in the event~\cite{Aaboud:2016urj}, which expresses the \pt projection of the neutrino pair of the invisibly decaying 
\PZ boson onto the \pt direction of the \PZ boson decaying to charged leptons, must exceed 50\GeV. The relative 
difference between \ptmiss and the dilepton \pt with respect to the dilepton \pt must be smaller than 0.3~\cite{Aaboud:2016urj}. Events with jets are rejected.

Most background contributions, including photon conversions, charge mismeasurement, and those processes yielding prompt leptons in the final state, 
such as \ttbar, single top, Drell--Yan (DY), and diboson production, are estimated from simulation. Backgrounds involving 
one or more nonprompt leptons are estimated from simulation aided by control samples in data in those categories with two leptons in the final 
state and exclusively from simulation otherwise.
Nonprompt-lepton background sources are composed of processes in which at least one of the final-state leptons does not come from the decay 
of a \PW or \PZ boson either directly or mediated by a leptonic \Pgt decay. Dominant SM sources of nonprompt background depend on the 
specific decay channel: \PZ{}+jets and dileptonic \ttbar production for those channels with at least three leptons, \PW{}+jets and 
semileptonic \ttbar for channels with two leptons in the final state. The nonprompt-lepton contribution is estimated using a lepton 
misidentification rate method~\cite{Sirunyan:2018shy} based on the misidentification rate measured in a simulated \ttbar sample and applied 
to control region data.
The main background contributions differ in each SR because of the different final states under study. The dominant background contributions 
for the \WW measurement come from nonprompt lepton and top quark production processes. For the \WZ 3$\ell$, the background is mainly arising from \PZ{}+jets. For the \WZ 2$\mu$ss and \ZZ 2$\ell$2$\nu$ SRs the main background is nonprompt leptons. The \ZZ 4$\ell$ SR is very clean with 
a small background contribution of nonprompt leptons.

Although the measurements presented in this Letter are dominated by the statistical limitation of the size of
the data set, the impact of different sources of systematic uncertainties is studied.

The lepton identification scale factors are estimated using the tag-and-probe method~\cite{Khachatryan:2010xn} as a function of the lepton 
\pt and $\eta$ for loose and tight electrons and muons. These scale factors are close to 1 and have an uncertainty 
of the order of 1--3\%, except for a few bins limited by statistics. Corrections to the jet energy scale are applied 
to data and simulation separately as $\eta$- and \pt-dependent corrections. Jet energy corrections are derived from simulation to bring 
the measured response of jets to that of particle-level jets on average. In situ measurements of the momentum balance in dijet, 
photon+jet, \PZ{}+jet, and multijet events are used to correct any residual differences in jet energy scale in data and 
simulation~\cite{Khachatryan:2016kdb}. The jet energy resolution amounts typically to 15\% at 10\GeV and 8\% at 100\GeV. During 
the 2017 data taking, a gradual shift in the input timing of the ECAL L1 trigger in the region of $\abs{\eta}>2.0$ 
caused a trigger inefficiency of 1\%--4\%. Correction factors were computed from data and applied to the acceptance evaluated by simulation.
For the selected events the trigger efficiency is very close to 1 in all channels, and no further correction is applied to the 
simulation. The relative difference between the trigger efficiency estimated in the \WW, \WZ, and \ZZ samples and 100\% is used 
as an uncertainty in the respective cross section measurement.  

The uncertainty arising from the choice of PDF is determined by reweighting the sample of simulated diboson events according to the 32 
replica PDF sets from PDF4LHC15~\cite{Butterworth:2015oua}. 
The envelope of the variations in the signal yields is used as the estimate of the uncertainty. The systematic bias 
due to the missing higher-order diagrams in \POWHEG is estimated by varying the default renormalization and factorization 
scale choices independently by a factor of 2 or 1/2. The uncertainty is assigned from the maximum 
difference in the signal yields for each variation, excluding the two extreme up or down combinations, with the nominal values. 

The uncertainty assigned to the nonprompt lepton contribution in the two-lepton categories is 
based on the differences between the lepton misidentification rate estimated for all flavor jets, and the \PQb-flavor jets 
and light-flavor jets, separately.
The uncertainty affects mostly electrons: a 30\% uncertainty is used per electron and 15\% per muon in the categories 
with two leptons in the final state. Normalization uncertainties of 30\% for conversions, 20\% for nonprompt leptons, charge 
mismeasurement, and diboson, and 10\% for top quark and DY processes are assigned.

An uncertainty of 1.9\% in the integrated luminosity, estimated offline using the methodology described in 
Ref.~\cite{CMS-PAS-LUM-19-001}, is applied as a global normalization uncertainty for all processes. 
 
The statistical uncertainties due to the limited size of the MC samples are treated according to the Barlow--Beeston method~\cite{Barlow:1993dm}; 
individual nuisance parameters (per process and per channel) are used when the corresponding expected amount of events in the bin is smaller than 10 events.

A summary of the expected event yields for signal and each of the background processes, and the observed data 
in the \WW SR is shown in Table~\ref{tab:WW}. For the other SRs, the expected event yields for signal and 
the total background, and the observed data, are shown in Table~\ref{tab:otherSRs}. 

\begin{table}[h]
\centering
\topcaption{\label{tab:WW} Expected event yields in the \WW SR and observed number of events. The uncertainties 
correspond to the statistical and systematic component, respectively.}
\begin{scotch}{cc}
Source             & Number of events \\ 
\hline
Top quark          &   9.0 $\pm$        0.1 $\pm$        1.1   \\ 
\WZ{}+\ZZ            &   5.6 $\pm$        1.0 $\pm$        1.1   \\ 
Drell--Yan         &   1.8 $\pm$        0.5 $\pm$        0.2   \\ 
Conversions        &   2.7 $\pm$        0.7 $\pm$        0.7   \\ 
Nonprompt $\ell$   &  11.2 $\pm$        1.3 $\pm$        3.4   \\ [\cmsTabSkip]                                                                        
Background         &  30.3 $\pm$        1.9 $\pm$        3.9   \\ 
\WW signal        &  55.2 $\pm$        0.3 $\pm$        1.8   \\ [\cmsTabSkip]                                                                        
Data               &  101   \\ 
\end{scotch}
\end{table}

\begin{table}[h]
\centering
\topcaption{\label{tab:otherSRs} Expected event yields for the signal and total background in the \WZ and \ZZ SRs, and observed number of events. The uncertainties correspond to the statistical and systematic component, respectively.}
\begin{scotch}{cccc}
          SR        & Background               & Signal                   & Data \\  \hline                                                                         
\WZ 3$\ell$        & 4.0 $\pm$ 0.6 $\pm$ 0.4  & 14.8 $\pm$ 0.1 $\pm$ 0.6 & 12 \\ 
\WZ 2$\mu$ss       & 0.6 $\pm$ 0.1 $\pm$ 0.1  &  3.2 $\pm$ 0.8 $\pm$ 0.2 &  4 \\
\ZZ 4$\ell$        & 0.5 $\pm$ 0.2 $\pm$ 0.1  &  2.5 $\pm$ 0.0 $\pm$ 0.1 &  3 \\ 
\ZZ 2$\ell$2$\nu$  & 4.8 $\pm$ 0.3 $\pm$ 0.7  &  3.8 $\pm$ 0.0 $\pm$ 0.2 & 12 \\
\end{scotch}
\end{table}

The cross sections are measured in regions, called total regions, defined to provide a measurement without any detector acceptance 
requirements. 
As outlined below, each channel has a different total region defined at generator level with dressed leptons (the momenta of generator-level 
photons within a cone of $\DR(\ell,\cPgg) <$ 0.1 is added to the lepton momenta).
All total regions are defined as excluding 
events containing any OSSF lepton pair with invariant mass below 4\GeV. For the \WW total region, no further requirements are applied.
For the \WZ total region, an additional kinematic requirement is imposed that selects events consistent with the on-shell \PZ boson production: 
$60 < m_{\lepZI, \lepZII} < 120\GeV$. For the \ZZ total region, this  additional kinematic requirement is applied to both \PZ boson candidates. 
The lepton tagging algorithm defined above to assign leptons to either the \PW or \PZ boson decay is applied. In the case of the 
\ZZ 2$\ell$2$\nu$ SR, one \PZ boson is reconstructed from the two leptons and the other from the neutrinos.

\begin{figure}[h]
  \centering
  \includegraphics[width=0.45\linewidth]{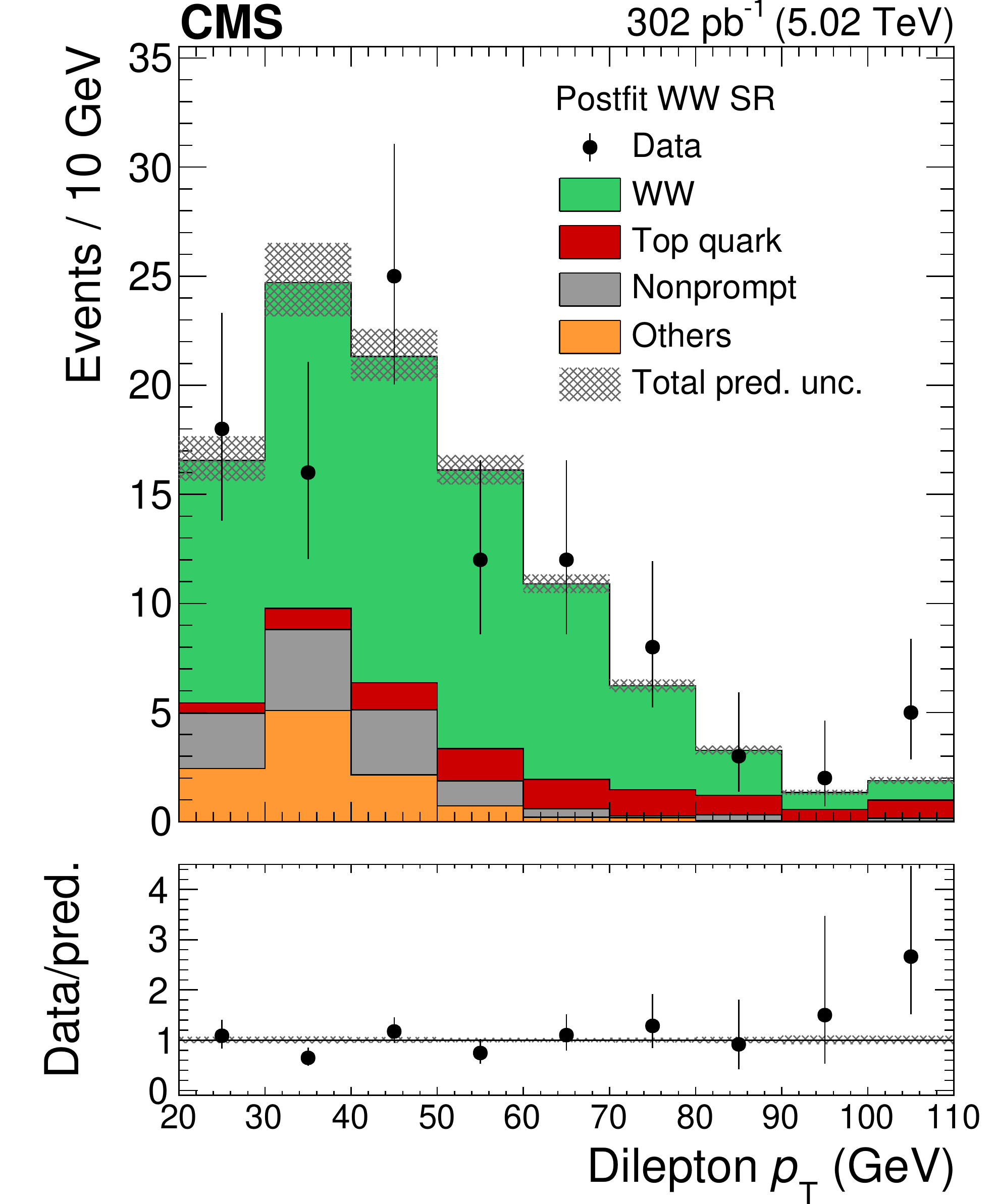}
  \includegraphics[width=0.45\linewidth]{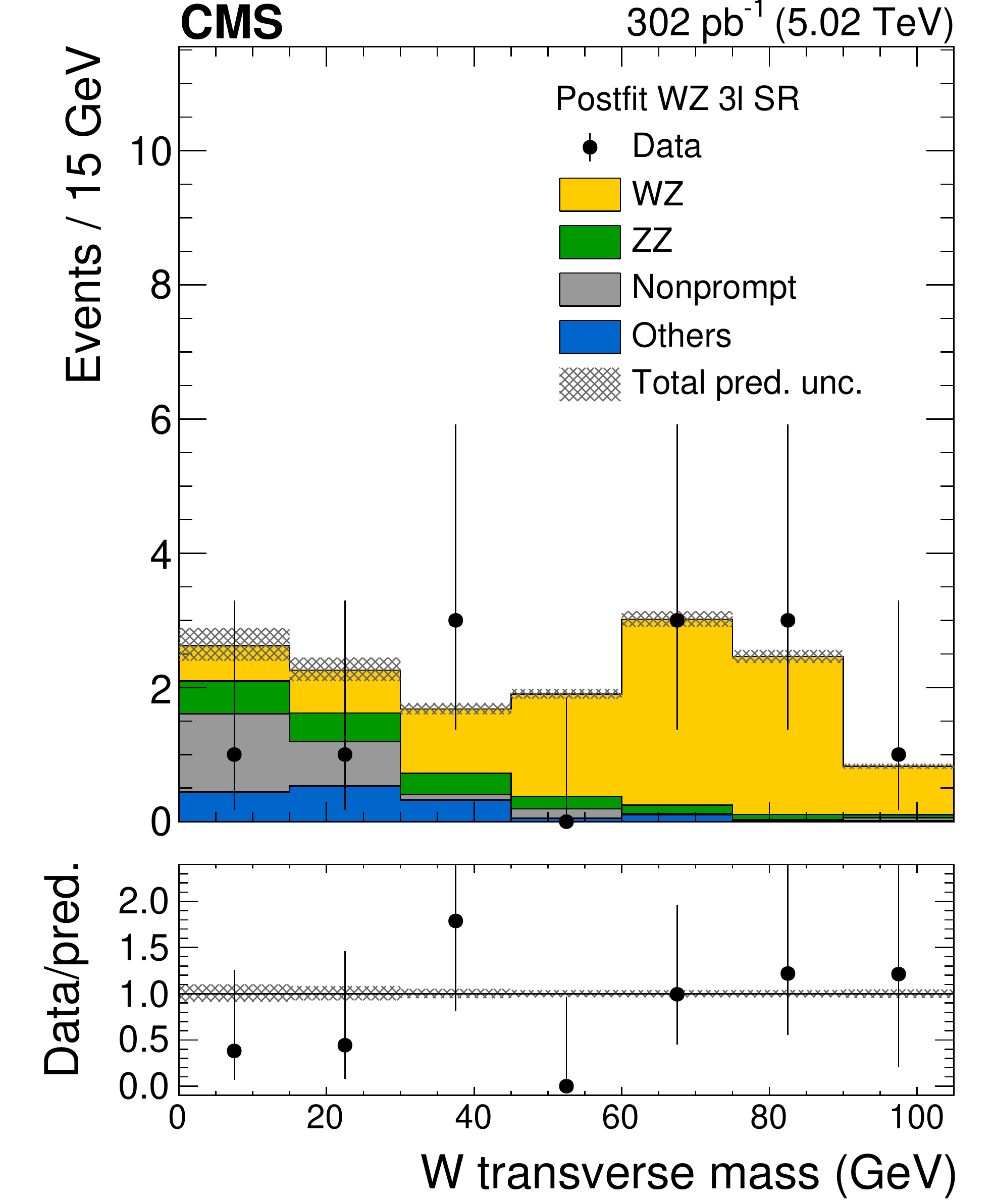}
  \caption{Distribution of the dilepton \pt in the \WW signal region (left). Events from DY, conversions, and diboson processes are grouped into the 'Others' category. Distribution of the \PW boson transverse mass in the \WZ 3$\ell$ signal region (right). Events from conversions, and DY processes are grouped into the Others category. The vertical error bars represent the statistical uncertainty in the data and the shaded band the uncertainty in the prediction. The signal contributions are scaled to the measured cross sections (postfit). }
  \label{fig:fig1}
\end{figure}

The total cross section, $\sigma$, is computed as
\begin{equation}
\label{eq:totxs}
\sigma = \frac{N_{\text{signal}}^{\mathrm{SR}}}{{\mathcal{B}}(\PW\to \ell\nu \text{ or } \PZ\to\ell\ell) \ {\mathcal{B}}(\PW\to \ell\nu \text{ or } \PZ\to\ell\ell) \ \epsilon \ \mathcal{L}},
\end{equation}
where $\mathcal{L}$ is the total integrated luminosity, $\epsilon$ is the efficiency of the lepton reconstruction 
and the additional phase space requirements, and $N_{\text{signal}}^{\mathrm{SR}}$ is the number of obtained signal events, 
estimated for each SR by performing a maximum likelihood fit to the yields with a single free-floating parameter 
that corresponds to the normalization of the signal process. The efficiency  values are computed using the signal 
simulated samples as the ratio of the number of events that fulfill the SR requirements over those that only pass 
the total region ones. An extrapolation from the lepton final state to the total production cross section is done 
by dividing by the branching fractions (${\mathcal{B}}$) of each of the \PW and/or \PZ bosons to leptons, which 
are taken from Ref.~\cite{Zyla:2020zbs}. The distributions of the dilepton \pt and \PW boson transverse mass in the 
\WW and \WZ 3$\ell$ signal region, respectively, are shown in Fig.~\ref{fig:fig1}. Good agreement between the 
observed data and prefit and postfit predictions is found in all channels.

\begin{figure}[h]
  \centering
  \includegraphics[width=0.9\linewidth]{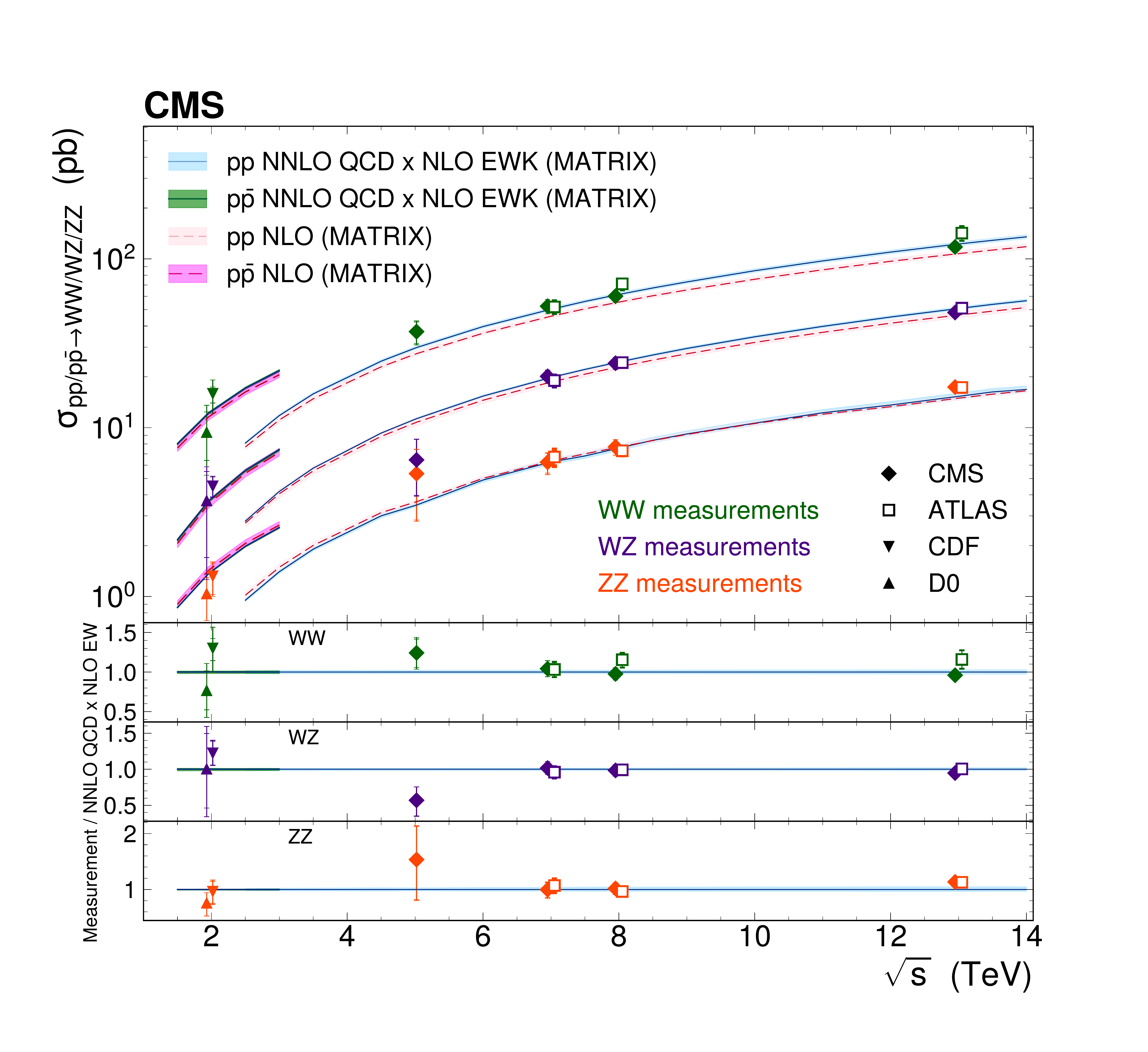}
  \caption{Results obtained in this analysis and other diboson production cross section measurements at 
different center-of-mass energies for the CMS~\cite{Chatrchyan:2013yaa, Khachatryan:2015sga, Sirunyan:2020jtq, 
Khachatryan:2016poo, Sirunyan:2019bez, Chatrchyan:2012sga, CMS:2014xja, Sirunyan:2020pub}, ATLAS~\cite{ATLAS:2012mec, 
Aad:2016wpd, Aaboud:2017qkn, Aad:2012twa, Aad:2016ett, Aaboud:2019gxl, Aad:2012awa, Aaboud:2016urj, Aaboud:2017rwm}, 
CDF~\cite{Aaltonen:2016pos, Aaltonen:2014yfa}, and D0~\cite{Abazov:2011cb, Abazov:2012cj, D0:2013rca} Collaborations 
are presented, and compared with the NNLO QCD $\times$ NLO EW and NLO predictions from \MATRIX. The vertical error 
bars represent the uncertainty in the measured cross section. }
  \label{fig:sigmaSumm}
\end{figure}

The uncertainties derived above are propagated to the final result through the numerator of Eq.~\eqref{eq:totxs}.
The measured values for the \WW, \WZ, and \ZZ total cross sections, shown in Fig.~\ref{fig:sigmaSumm}, are
\begin{equation*}
\label{eq:sigmaww}
\begin{split}
  \sigma_{\WW}=37.0^{+5.5}_{-5.2}\stat^{+2.7}_{-2.6}\syst & =37.0^{+6.2}_{-5.8}\unit{pb},\\
  \sigma_{\WZ}=6.4^{+2.5}_{-2.1}\stat^{+0.5}_{-0.3}\syst & =6.4^{+2.5}_{-2.1}\unit{pb},\\
  \sigma_{\ZZ}=5.3^{+2.5}_{-2.1}\stat^{+0.5}_{-0.4}\syst & =5.3^{+2.6}_{-2.1}\unit{pb},
\end{split}
\end{equation*}
respectively. Figure~\ref{fig:sigmaSumm} also presents a summary of the 
diboson production cross section measurements at different center-of-mass energies and a comparison with fixed-order 
predictions produced via the \MATRIX\ framework~\cite{Grazzini:2017mhc}. For the \WZ measurement, the result is 
consistent with the SM prediction within two standard deviations. The calculations are performed with 
the NNPDF31$\_$nnlo$\_$as$\_$0118$\_$luxqed~\cite{Bertone:2017bme} PDF set (NNPDF31$\_$nnlo$\_$as$\_$0118$\_$luxqed$\_$nf4 
for \WW production). The quark-induced processes are calculated at NNLO in QCD and NLO in electroweak (EW) corrections. For the \WW and 
\ZZ processes, the gluon induced contribution is calculated at NLO in QCD~\cite{Grazzini:2018owa}. Photon-induced 
contributions are included at up to NLO EW. The quark-induced NNLO QCD and NLO EW contributions are combined 
multiplicatively (NNLO QCD $\times$ NLO EW), and the gluon- and photon-induced contributions are combined additively, following the procedure 
described in Ref.~\cite{Kallweit:2019zez}.

The diboson production cross sections are measured for the first time at a new energy, 5.02\TeV, using data collected 
with the CMS detector corresponding to an integrated luminosity of 302\pbinv. The analysis is performed in the leptonic 
decays of the \PW and \PZ bosons with at least two leptons in the final state. The measured total cross sections are 
consistent with theoretical calculations at combined next-to-next-to-leading order quantum chromodynamics and next-to-leading 
order electroweak accuracy.

\begin{acknowledgments}
  We congratulate our colleagues in the CERN accelerator departments for the excellent performance of the LHC and thank the technical and administrative staffs at CERN and at other CMS institutes for their contributions to the success of the CMS effort. In addition, we gratefully acknowledge the computing centers and personnel of the Worldwide LHC Computing Grid and other centers for delivering so effectively the computing infrastructure essential to our analyses. Finally, we acknowledge the enduring support for the construction and operation of the LHC, the CMS detector, and the supporting computing infrastructure provided by the following funding agencies: BMBWF and FWF (Austria); FNRS and FWO (Belgium); CNPq, CAPES, FAPERJ, FAPERGS, and FAPESP (Brazil); MES (Bulgaria); CERN; CAS, MoST, and NSFC (China); MINCIENCIAS (Colombia); MSES and CSF (Croatia); RIF (Cyprus); SENESCYT (Ecuador); MoER, ERC PUT and ERDF (Estonia); Academy of Finland, MEC, and HIP (Finland); CEA and CNRS/IN2P3 (France); BMBF, DFG, and HGF (Germany); GSRT (Greece); NKFIA (Hungary); DAE and DST (India); IPM (Iran); SFI (Ireland); INFN (Italy); MSIP and NRF (Republic of Korea); MES (Latvia); LAS (Lithuania); MOE and UM (Malaysia); BUAP, CINVESTAV, CONACYT, LNS, SEP, and UASLP-FAI (Mexico); MOS (Montenegro); MBIE (New Zealand); PAEC (Pakistan); MSHE and NSC (Poland); FCT (Portugal); JINR (Dubna); MON, RosAtom, RAS, RFBR, and NRC KI (Russia); MESTD (Serbia); SEIDI, CPAN, PCTI, and FEDER (Spain); MOSTR (Sri Lanka); Swiss Funding Agencies (Switzerland); MST (Taipei); ThEPCenter, IPST, STAR, and NSTDA (Thailand); TUBITAK and TAEK (Turkey); NASU (Ukraine); STFC (United Kingdom); DOE and NSF (USA).
\end{acknowledgments}

\bibliography{auto_generated}   

\cleardoublepage \appendix\section{The CMS Collaboration \label{app:collab}}\begin{sloppypar}\hyphenpenalty=5000\widowpenalty=500\clubpenalty=5000\vskip\cmsinstskip
\textbf{Yerevan Physics Institute, Yerevan, Armenia}\\*[0pt]
A.~Tumasyan
\vskip\cmsinstskip
\textbf{Institut f\"{u}r Hochenergiephysik, Wien, Austria}\\*[0pt]
W.~Adam, J.W.~Andrejkovic, T.~Bergauer, S.~Chatterjee, M.~Dragicevic, A.~Escalante~Del~Valle, R.~Fr\"{u}hwirth\cmsAuthorMark{1}, M.~Jeitler\cmsAuthorMark{1}, N.~Krammer, L.~Lechner, D.~Liko, I.~Mikulec, P.~Paulitsch, F.M.~Pitters, J.~Schieck\cmsAuthorMark{1}, R.~Sch\"{o}fbeck, M.~Spanring, S.~Templ, W.~Waltenberger, C.-E.~Wulz\cmsAuthorMark{1}
% ...
\vskip\cmsinstskip
% ...
1:  Also at TU Wien, Wien, Austria\\
% ...
\end{sloppypar}
\end{document}